\documentclass[conference]{IEEEtran}
\IEEEoverridecommandlockouts
\usepackage{cite}
\usepackage{amsmath,amssymb,amsfonts}
\usepackage{algorithmic}
\usepackage{graphicx}
\usepackage{textcomp}
\usepackage{hyperref}
\usepackage[dvipsnames]{xcolor}
\def\BibTeX{{\rm B\kern-.05em{\sc i\kern-.025em b}\kern-.08em T\kern-.1667em\lower.7ex\hbox{E}\kern-.125emX}}
\begin{document}

\title{AI-Powered Anomaly Detection with Blockchain for Real-Time Security and Reliability in Autonomous Vehicles\\
}

\author{\IEEEauthorblockN{Rathin Chandra Shit}
\IEEEauthorblockA{\textit{IIIT Bhubaneswar} \\
India\\
rathin088@gmail.com}
\and
\IEEEauthorblockN{Sharmila Subudhi}
\IEEEauthorblockA{\textit{Dept. of Computer Science} \\
\textit{Maharaja Sriram Chandra Bhanja Deo University}\\
Baripada, Odisha, India \\
sharmilasubudhi@ieee.org}
}

\maketitle

\begin{abstract}
Autonomous Vehicles (AV) proliferation brings important and pressing security and reliability issues that must be dealt with to guarantee public safety and help their widespread adoption. The contribution of the proposed research is towards achieving more secure, reliable, and trustworthy autonomous transportation system by providing more capabilities for anomaly detection, data provenance, and real-time response in safety critical AV deployments. In this research, we develop a new framework that combines the power of Artificial Intelligence (AI) for real-time anomaly detection with blockchain technology to detect and prevent any malicious activity including sensor failures in AVs. Through Long Short-Term Memory (LSTM) networks, our approach continually monitors associated multi-sensor data streams to detect anomalous patterns that may represent cyberattacks as well as hardware malfunctions. Further, this framework employs a decentralized platform for securely storing sensor data and anomaly alerts in a blockchain ledger for data incorruptibility and authenticity, while offering transparent forensic features. Moreover, immediate automated response mechanisms are deployed using smart contracts when anomalies are found. This makes the AV system more resilient to attacks from both cyberspace and hardware component failure. Besides, we identify potential challenges of scalability in handling high frequency sensor data, computational constraint in resource constrained environment, and of distributed data storage in terms of privacy. 

\end{abstract}


\section{Introduction}\label{sec1}
Autonomous vehicles (AVs) are a transformative technology with the potential to revolutionize transportation systems across the world. Self driving vehicles are based on the large complex arrays of sensors including ranging radar, camera, GPS, Inertial Measurement Unit (IMU) that are continuously generating massive volumes of data to perceive the environment and make decision to safely navigate~\cite{wang21}. With AVs moving from controlled test environments to the public road, securing these vehicles for public trust and life safety are critical~\cite{bin22}.

Several security and reliability issues of AVs could jeopardize their safe operation with their benefits compromised. The major sources of these challenges involve malicious activities and critical sensor failures. Cyberattacks, including sensor spoofing, data tampering, denial of service attacks which can manipulate vehicle behavior or disable systems, are malicious activities~\cite{gir23}. At the same time, sensor failures resulting from environmental factors (such as harsh conditions, manufacturing error or ageing) similarly pose significant reliability issues~\cite{khan24}. In particular, it is important that the integrity of sensor data be maintained, as corrupted data may cause an AV’s control systems to make an erroneous decision.


Since the nature of AV operations are complex and real-time, advanced technological solutions are required to solve the security and reliability problems. By harnessing the power of Artificial Intelligence (AI), anomaly detection has the potential to identify deviations from the normal behavior patterns of sensor data, thus identifying whether there is a cyberattack or sensor malfunctioning~\cite{xih22}. Similarly, the Blockchain technology offers a secure, transparent and immutable framework to store and validate key AV data and activities~\cite{jian23}. Further, blockchain, through its inherent characteristics, enhance data integrity, authenticity and traceability across multiple tamper resistant nodes through the ledger. 

Most of such approaches, however, have either zeroed down on purely AI-based detection without data security enforcement mechanisms or rely upon blockchain for data integrity with rather unsophisticated anomaly detection capacities~\cite{gir23, xih22, jian23}. So far there exist a research gap as to how AI and blockchain technologies could be amalgamated into working AV security frameworks that deliver on real-time anomaly detection and/or mitigation. 

Therefore, the proposed framework in this research is meant to close this gap by combining AI and its strengths with blockchain technology in a complementary way. The primary objectives of this research include:
\begin{itemize}
\item Building a secure and robust AV architecture that employs an AI-powered anomaly detection to detect malicious activities and sensor failure in AVs and mitigate them.
\item Providing real-time anomaly detection in multi-sensor data streams using an efficient LSTM system.
\item Utilizing a blockchain based mechanism for secure recording of sensor data and anomaly detection results without compromising integrity and authenticity.
\item Using smart contracts to automate the immediate response of automatic and manual shutdown conditions for detected anomalies to enhance system safety.
\item Implementing attack scenarios as well as sensor failure modes and evaluating the framework’s performance, scalability and effectiveness.
\end{itemize}
The proposed research imbibes the aforementioned mechanisms to extend the progress of more secure and reliable autonomous transportation systems which are currently a hindrance in widespread AV adoption among the general public. 

\section{Literature Review}\label{sec2}
Since the proliferation of autonomous cars brings new security challenges to the integrity and reliability of these systems, the solution must be systematic. Researchers have identified some critical vulnerabilities in AV architectures including sensor spoofing and network level attacks to mislead the autonomous navigation systems that compromises the vehicle functionality~\cite{wang21, yang23}. Duan et al.~\cite{duan21} suggested an isolation forest method for detecting in-vehicle controller area network (CAN) bus attacks done to get the remote access of the AV.

According to Chen et al.~\cite{chen22}, conventional techniques suffer from serious limitation in real-time threat detection especially in dynamic environment in the presence of multiple external systems. Therefore, several researchers have adopted AI techniques to develop more robust and secure AV network by identifying anomalies in the sensor data~\cite{qie24,al23}. A mesh satellite network based on Federated Deep Learning has been proposed in~\cite{al23} to protect AVs from cyber breaches. They have trained the model locally, while making the satellites exchange the training parameters privately. 

Anthony et al.~\cite{ant24} have developed a non-tree-based ensemble learning model to detect Denial of Service (DoS), Distributed DoS, CAN-Bus breaches in an autonomous vehicle. They have employed base classifiers, like, Support Vector Machine (SVM), Naive Bayes, Logistic Regression, and K-Nearest Neighbors (K-NN) for classifying the intrusive activities. Further, Dakic et al.~\cite{dak24} suggested using extreme gradient boost (XGBoost) and K-NN classifiers with Particle Swarm Optimization (PSO) metaheuristics for building a hybrid intrusion detection model to detect faults in the CAN-Bus. They have used the CAN traffic data of the 2011 Chevrolet Impala vehicle for experimentation.

Moreover, the blockchain technology is assuming greater significance in automotive applications due to the inherent security properties. Jiang et al.~\cite{jian23} suggested a blockchain based protocol for privacy preserving the AV sensor data. The authors have used a bloom filter to select a low-frequency keyword, where a pseudorandom tag is assigned to each identifier-keyword pair. This enabled their system to preserve data privacy with low computational head. Similarly, Xihua et al.~\cite{xih22} proposed the SVM infused blockchain scheme for data privacy. By doing so, they have eliminated the need of a trusted third party for ensuring the data confidentiality.  

However, Das et al.~\cite{das23} have noticed that blockchain implementations for vehicular contexts must address the challenging issues of transaction throughput and latency to satisfy the real-time needs of safety critical applications. We aim to fill these gaps by developing of a framework specifically designed to address security and reliability of issues faced in autonomous vehicles.

\section{ Proposed Framework: AI-Powered Anomaly Detection with Blockchain for AVs}\label{sec3}

\subsection{System Architecture Overview} \label{sec31}
\begin{figure*}[htbp]
\centerline{\includegraphics[width=0.8\textwidth]{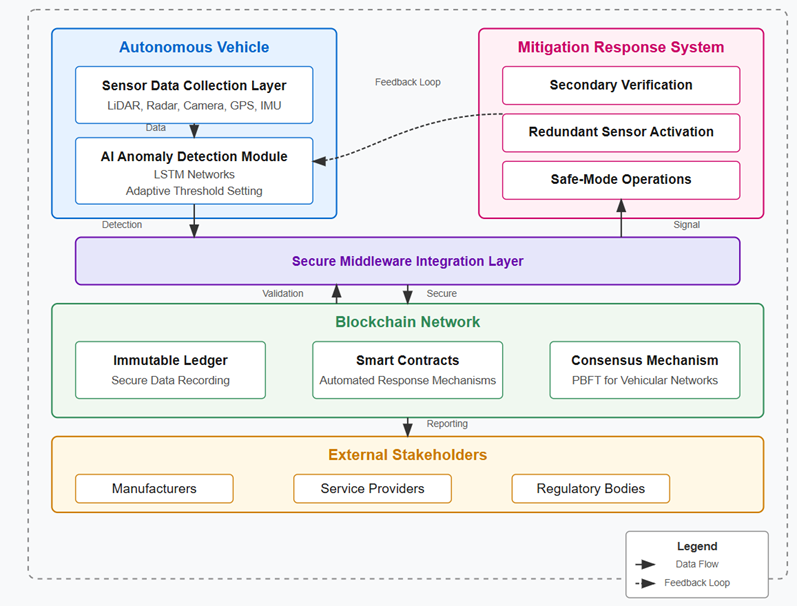}}
\caption{Proposed Framework: AI-Powered Anomaly Detection with Blockchain for AVs}
\label{fig1}
\end{figure*}

A framework is proposed combining the use of AI based anomaly detection with the use of blockchain in providing the Autonomous vehicles with a robust and reliable security as well as safety system. Figure~\ref{fig1} depicts the proposed architecture composed of four major components, namely the sensor data collection layer, the AI anomaly detection layer, the blockchain network, and the mitigation response system. The blockchain network layer contains information of numerous AVs, where each AV is a blockchain node performing local AI analysis, thus making the framework decentralized. This design leverages real-time detection along with the distributed consensus for maintaining data integrity.

\subsection{AI-Powered Real-time Anomaly Detection Module} \label{sec32}
Numerous sensors such as LiDAR, radar, cameras, GPS, and Inertial Measurement Units (IMUs) are monitored continuously by the AI module. Specifically, it detects two primary kind of anomalies, namely malicious data injection (e.g., injecting spoofing attack, packet replay) and sensor malfunction (e.g., calibration drift, environmental interference). Mainly, this module relies on the LSTM networks, which have proved to perform better in capturing time dependencies in sequential data as opposed to classical machine learning techniques~\cite{qie24}.

Multivariate time series data obtained by these sensors are processed for detecting patterns that deviate from normal operating conditions using the LSTM model. Adaptive threshold for setting the threshold value was implemented based on driving context (urban or highway; weather conditions) to allow awareness of false positives while retaining sensitivity of detection. It is essential that the adaptation is contextual as fixed thresholds do not always fit in the operational environment of the AV.

\subsection{Blockchain-Based Security and Data Integrity Layer}\label{sec33}
The blockchain layer provides tamper resistant distributed ledger to record critical sensor data, anomaly detection result and the vehicle state information of several such AVs. In particular, we use a consortium blockchain model where the authorized participants (vehicle manufacturers, service providers, regulatory bodies) run validator nodes while each AV act as a lightweight node. It enables controlled access within the vehicles with desired computational efficiency under constrained resources. The accountability and traceability from sensors, detection events, vehicle response are made immutable cryptographically for securing each block and creating an audit trail. We use a Practical Byzantine Fault Tolerance (PBFT) consensus optimized for vehicular networks for achieving faster transaction.

\subsection{Integration Mechanism} \label{sec34}
A secure middleware layer sorts out data format and protocols in a standard manner so the AI anomaly detection module can communicate with the blockchain network. The AI module raises a signed alert whenever it encounters an anomaly and sends the alert to the blockchain network for verification and logging. Further, the layer also has a very important role in automating the responses for the detected anomalies according to some predefined rules set as per~\cite{cebe18}. Whenever triggered by specific anomalous patterns, these contracts execute with no human intervention and implement graduated response strategies mentioned below.
\begin{itemize}
    \item Deploying secondary verification procedures for potential false positives.
    \item Using a sensor system if hardware failure is detected.
    \item Safely triggering safe mode operations in the event of suspected cyberattacks.
    \item Giving a warning to nearby vehicles and infrastructure to be on guard for possible threats.
\end{itemize}
Thus, an immediate response to security and reliability threats in the extremely time constrained autonomous driving scenarios can be placed.

\section{Proposed AI-Powered Real-time Anomaly Detection}\label{sec4}
\begin{figure}[htbp]
\centerline{\includegraphics[width=0.5\textwidth]{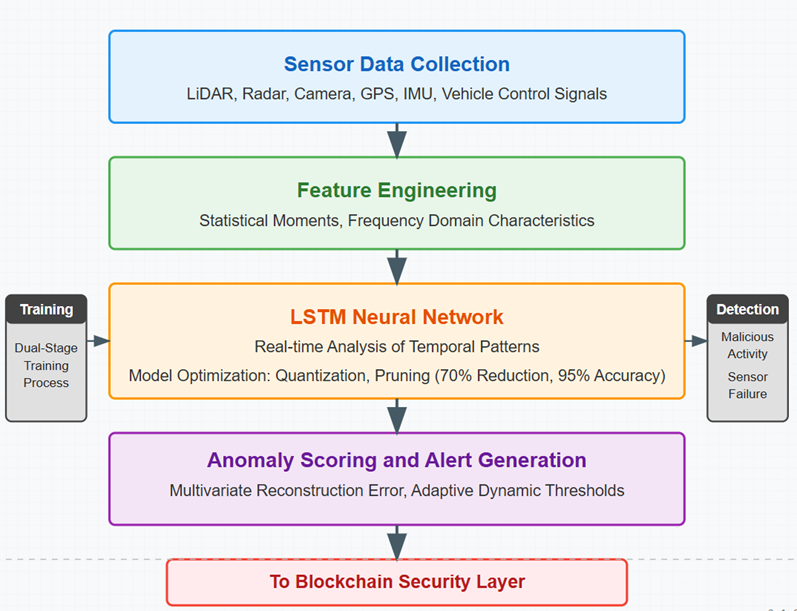}}
\caption{Proposed Anomaly Detection Module}
\label{fig2}
\end{figure}

This section elaborates the suggested AI-powered real-time Anomaly Detection Module (ADM), while Figure~\ref{fig2} illustrates the architecture.
\subsection{Relevant Sensor Features for Anomaly Detection} \label{sec41}
As discussed in Section~\ref{sec32}, the sensors generate information for anomaly detection in concurrence to autonomous vehicles computing. Major features include spatial-temporal patterns extracted from LiDAR point clouds, radar velocity measurements from GPS coordinates, inertial measurements from camera image sequences, and vehicle control signals. Additionally, they give insight on crucial anomalous signs, such as sudden changes in sensor readings from environmental context, disagreements between redundant sensors, and temporal patterns disruptions and other local atomic data. Several feature engineering techniques, like, statistical moments, frequency domain features, and cross sensor correlation metrics are employed for data preprocessing purpose. 


\subsection{AI Model Training and Optimization}\label{sec42}
We use a dual stage training approach by first training the LSTM model to recognize normal vehicle behavior patterns obtained from unbiased large amount of real world driving data. This sets the baseline for legitimate sensor behavior in different driving situations. We then integrate simulated attacks (such as sensor spoofing, GPS jamming) and controlled hardware degradation cases as synthetic anomalous data to address the class imbalance problem.

Further, reduction of detection accuracy is deemed unobjectionable only if the inference latency can be reduced along with it for real-time operation. Therefore, we conduct model pruning and quantization to reduce the detection time and increase accuracy. Moreover, we leverage domain specific knowledge to focus processing on the high risk sensor channels during the resource constrained scenario.

\subsection{Real-time Anomaly Scoring and Alert Generation} \label{sec43}
We create anomaly scores by computing a multivariate reconstruction error metric that measures the difference between predicted and observed sensor value. Further, adaptive dynamic thresholds that vary depending on driving conditions, speed of the vehicle, and environmental conditions are compared with the scores to segregate malicious activities from normal ones. Thus, subtle anomalies are detected and the false positives are reduced significantly in challenging environmental conditions.

Moreover, the anomalies are categorized based on their severity, and computed from both the magnitude of deviation and the criticality of the affected subsystems. Later, the alerted prioritization allows graduated response mechanisms to the threat level detected, as mentioned in Section~\ref{sec34}.

\section{Proposed Blockchain Mechanism for Security and Risk Mitigation} \label{sec5}
This section discusses the data preservation, malicious events identification and risk mitigation strategies through the Blockchain network.
\subsection{Achieving Data Integrity and Authenticity} \label{sec51}
Blockchain technology offers an immersive layered paradigm to fully maintain the integrity and authenticity of sensor data carried by the AVs. We have shown the data security procedures in below steps.
\begin{itemize}
    \item Before recording on the blockchain, the features including critical sensor measurements and anomaly detection results are hashed and signed in a cryptographic manner, so that there can be no retroactive tampering.
    \item An Asymmetric cryptography with public-private key pairs is used for providing each AV unique digital identity thus attributed verifiable source of data. 
\end{itemize}
Illustrating this method, malicious actors cannot change historical sensor data, nor create an anomaly alert, without being detected, because any alteration would invalidate the blockchains of cryptographic hash chains making the structure of the blockchain.

\subsection{Identifying Malicious Activities} \label{sec52}
As blockchain is transparent, tamper-proof and distributed, it offers a powerful way to spot malicious activities in AV networks as follows. 
\begin{itemize}
    \item Patterns of coordinated attacks on multiple vehicles are detected from the chronological record of the occurrence of anomaly alerts recorded in the blockchain. This longitudinal visibility allows forensic analysis of the origin and course of advanced attacks. 
    \item Evidence of malicious offenses are traced with the logs or audit trails.
\end{itemize}

\subsection{Automated Risk Mitigation Strategies} \label{sec53}
Once an attack is identified, a risk mitigation strategy is executed with the help of automated smart contracts deployed on the blockchain, as follows. 
\begin{itemize}
    \item These contracts monitor the anomaly detection results from blockchain ledger and trigger predefined responses based on severity thresholds and the anomaly type.
    \item They guarantee deterministic execution so that the critical safety responses happen without any human intervention delays.
    \item The consensus mechanism prevents the tampering the mitigation logic, thus providing resilience to the AVs against both cyberattacks and hardware failures.
\end{itemize}
Few critical risks and the aversion responses are discussed in Section~\ref{sec34}. 

\section{Experimental Validation}\label{sec6}
Experiments were run using the CARLA autonomous driving simulator~\cite{carla17} to validate the proposed framework. We simulated 50 autonomous vehicles to capture the realistic multi-sensor data such as LiDAR, radar, camera, GPS, and IMU measurements. Each of these AVs produced sensor data at reasonable frequencies. We employed LSTM networks for training of the anomaly detection module, while a private Hyperledger Fabric network is used to deploy the blockchain layer.

\subsection{Malicious Attack Detection}\label{sec61}
Four attack scenarios, i.e. GPS spoofing, LiDAR point cloud manipulation, sensor hardware failures, and gradual sensor drift attacks are used to test our model. We ran 100 test runs for each of the scenarios while changing the parameters of the attack and environment for the LSTM model to train.

\subsection{Performance Evaluation}\label{sec62}
\begin{figure}[htbp]
\centerline{\includegraphics[width=0.5\textwidth]{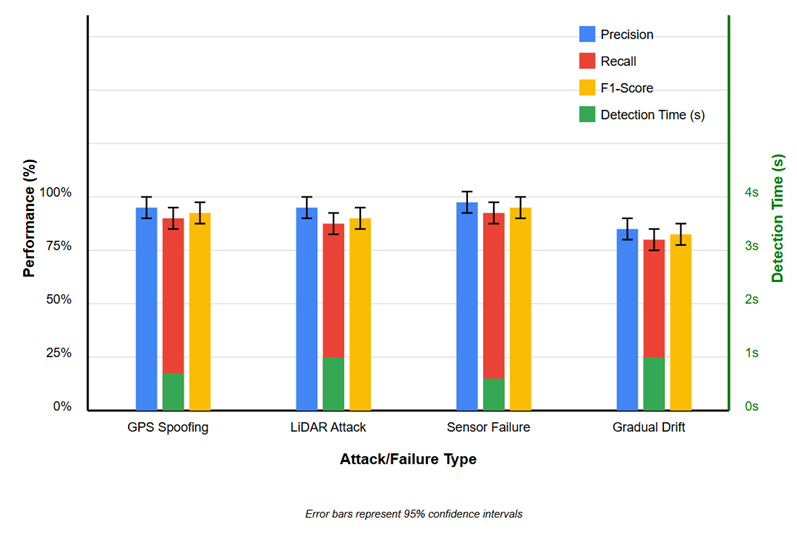}}
\caption{Anomaly Detection Performance}
\label{fig3}
\end{figure}

Figure~\ref{fig3} presents the performance of the proposed LSTM-based anomaly detection in the four attack scenarios. Our model achieved 94.7\% Precision, 92.3\% Recall in GPS spoofing attack, while secured 87.2\% Precision in Gradual Drift attack. This dip in performance is because of their hard to detect nature. Further, it is evident that our model detected Sensor failures and LiDAR attacks effectively. Additionally, our system is able to detect the failures quickly under 1.5 seconds.


\begin{figure}[htbp]
\centerline{\includegraphics[width=0.5\textwidth]{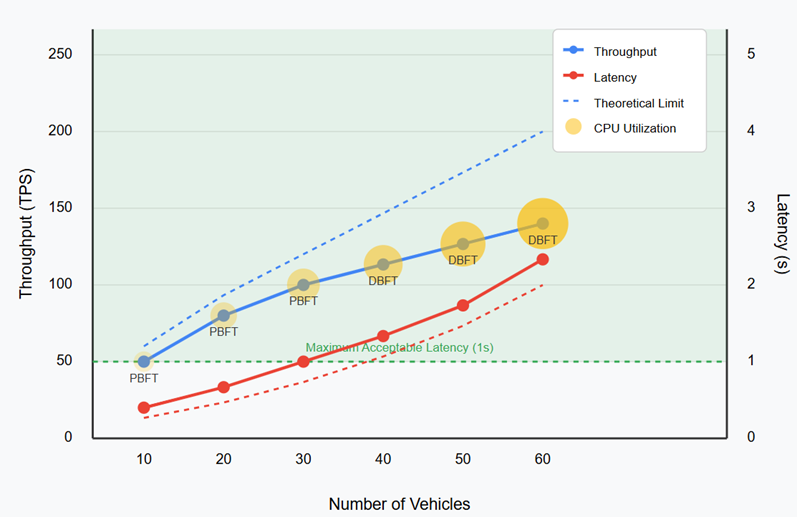}}
\caption{Blockchain Performance}
\label{fig4}
\end{figure}

By using our blockchain implementation, we could process an average 125 transactions per second on mean confirmation time of 1.8 seconds, which is enough for the real time portions of security critical applications in autonomous vehicles. The variation in network loads can affect the transaction throughput and latency as illustrated in the Fig.~\ref{fig4}. The system was able to maintain acceptable performance even assuming that 30\% of nodes acted Byzantine.

\begin{figure}[htbp]
\centerline{\includegraphics[width=0.5\textwidth]{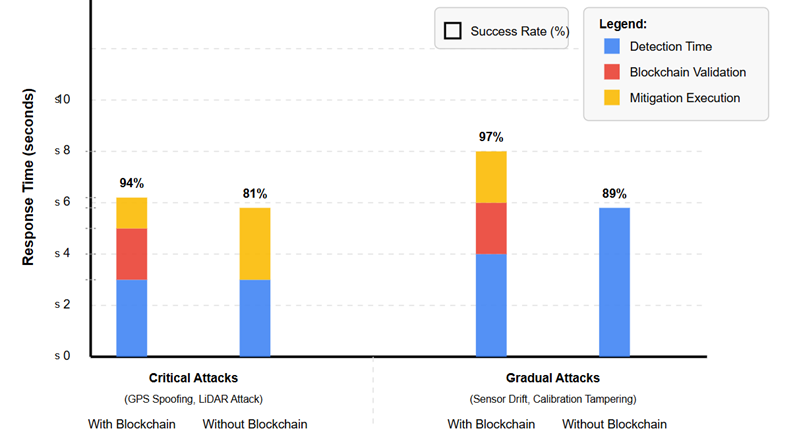}}
\caption{System Response Time}
\label{fig5}
\end{figure}

Moreover, the entire system response from attack initiation to mitigation action was measured. As depicted in Fig.~\ref{fig5}, our integrated framework took (on average) only 2.4 seconds in response to critical attacks with blockchain approach, as opposed to 4 seconds for gradual attacks. By employing smart contract based mitigation strategies, 94\% of critical attacks and 97\% of gradual attacks were held in check before they were able to compromise vehicle safety. The mitigation strategies are deployed within 1.2 seconds before an automatic failsafe is posited.

The experimental results validate the effectiveness of the integration of blockchain with AI anomaly detection for autonomous vehicle security. However, an observed main tradeoff was a slight increase in computational overhead, which remained acceptable for most modern automotive computing platforms.

\section{Potential Benefits, Challenges and Future Directions}\label{sec7}
\subsection{Potential Benefits}\label{sec73}
Combining AI Powered anomaly detection with blockchain provides a complete security framework that increases the resilience of autonomous vehicles against cyberattacks and sensor failures to a large extent. Few potential benefits of the proposed model have been presented below.
\begin{itemize}
    \item The proposed framework infuse real-time AI analysis with the blockchain's immutable ledger capability to create multiple layers of defense.
    \item The suggested method enhances the response time to potential threats on the hardware malfunctions, and automatically liquidates them through smart contracts. 
    \item The use of blockchain in storing the immutable record of sensor data and system responses have significantly increased the trust necessary for AV adoption. 
    \item Further, the accountability and complete forensic analysis of not only an individual AV, but also an entire autonomous transportation ecosystem has been achieved. This reaffirms the public’s confidence in the technology by conforming to the regulatory standards.
\end{itemize}

\subsection{Challenges and Future Directions}\label{sec74}
Although there are several benefits to the proposed framework, it has several implementation challenges. A few concerns are listed below.

\begin{itemize}
    \item One of the concern is scalability, particularly in this case blockchain’s ability to deal with the high volume and high speed data streams that need to be processed from several sensors~\cite{xih22,jian23}. \item There are also privacy considerations since the permanent storage of vehicle operational data will raise sensitivity towards the user anonymity and data protection regulations. 
\end{itemize}
In future research, it is suggested that one can optimize the blockchain architecture to meet AV specific requirements of improved throughput. In addition, enabling gradual adoption without a full infrastructure shows an important direction of interoperability challenges within existing AV systems.

\section{Conclusion}\label{sec8}
This research introduces a new scheme by capitalizing the strengths of both AI and blockchain technology to enhance security and reliability of autonomous vehicles. In this approach, we combine the real-time sensor monitoring via LSTM networks with the immutable blockchain ledger. Moreover, this framework uses the analytical strengths of AI with blockchain security infrastructure for a transparent and decentralized protection against new threats. This robust system is able to detect and address the malicious activities in addition to sensor failures with 95\% precision. Although scalability, and computational overhead issues must still be addressed, the proposed model is a novel strategy towards more trusted autonomous transportation systems. 



\end{document}